\documentclass[journal]{IEEEtran}
\ifCLASSINFOpdf
  \usepackage[pdftex]{graphicx}
\else
\fi
\usepackage{tikz}

\usepackage{comment}
\newcommand{\pluseq}{\mathrel{+}=}

\begin{document}
%
\title{Open-Source Design of Heterogeneous SoCs for AI Acceleration: the PULP Platform Experience}

\title{Open-Source Heterogeneous SoCs for AI\\ \Large The PULP Platform Experience}

%
%
%

\author{Francesco~Conti,~\IEEEmembership{Senior~Member,~IEEE,}
        Angelo~Garofalo,~\IEEEmembership{Member,~IEEE,}\\
        Davide~Rossi,~\IEEEmembership{Senior~Member,~IEEE,}
        Giuseppe~Tagliavini,~\IEEEmembership{Member,~IEEE,}
        and~Luca~Benini,~\IEEEmembership{Fellow,~IEEE}
\thanks{F.~Conti, A.~Garofalo, D.~Rossi and L.~Benini are with the Department
of Electrical, Electronic and Information Engineering, University of Bologna, 40123 Bologna, Italy, e-mail: \{f.conti,davide.rossi,luca.benini\}@unibo.it.}
\thanks{G.~Tagliavini is with the Department
of Computer Science and Engineering, University of Bologna, 40123 Bologna, Italy, e-mail: giuseppe.tagliavini@unibo.it.}
\thanks{A.~Garofalo and L.~Benini are also with the Integrated Systems Laboratory, ETH Z\"urich, 8005 Zurich, Switzerland.}
\thanks{Preprint submitted to Solid-State Circuits Magazine in December 2024.}}

\maketitle



%
\IEEEpeerreviewmaketitle

\IEEEPARstart{T}{he} complexity of Artificial Intelligence (AI) algorithms increases at an exponential pace that pure technological scaling, especially with the slowing of Moore's law, can not keep up with.
Epoch~AI estimates that the number of parameters in AI models is currently (as of 2024) scaling at a rate of 2$\times$ per year; training floating-point operations (FLOPs) are scaling even faster at 4.2$\times$ per year~\cite{EpochNotableModels2024}.
On the other hand, the same institution estimates that compute performance from dedicated hardware only scales at a rate of 1.3$\times$ per year for 32-bit floating-point data, and with similar rates for other data formats~\cite{EpochMachineLearningHardware2024}.
This figure includes gains from both technology node advancements and architectural improvements.


This setup creates an extraordinary challenge for the designers of heterogeneous AI System-on-Chips (SoCs).
On the one hand, accelerator designs must scale continuously to match the increasing complexity of AI workloads -- and this is true not only for datacenter AI accelerators but also for edge AI devices, whose functionality is expected to become progressively more sophisticated.
On the other hand, this scaling also needs to happen at a fast pace, which makes it imperative to design, verify, and tape-out new complex heterogeneous SoCs with a much quicker turnaround time than in traditional cycles -- especially for fabless startups.

\begin{figure}[tb]
    \centering
    \includegraphics[width=0.95\linewidth]{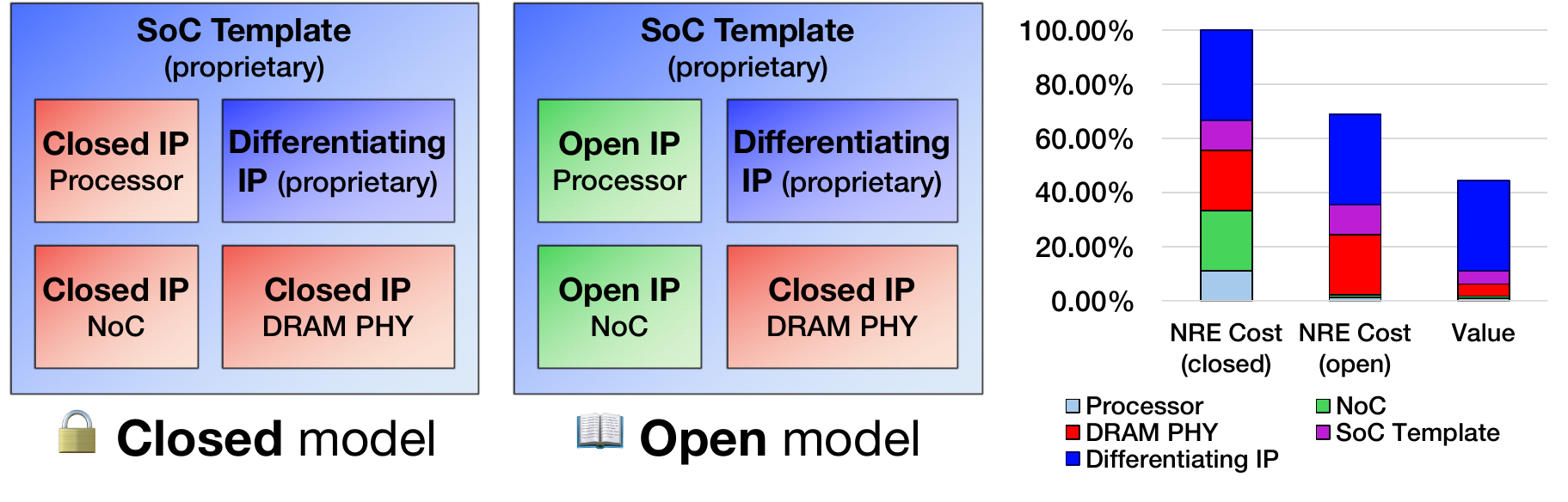}
    \caption{Example of the difference between a model of SoC design based on closed-source IP and one exploiting an open model. Exploiting an open-source model lowers the non-recurrent engineering costs related to the design of IPs that are not associated with most of a SoC's value, such as key proprietary IPs developed by a startup, lowering the access barriers and freeing up funding for the development of the high-value proprietary IPs.}
    \label{fig:open_model}
\end{figure}

By merit of its ``automatic'' cost-sharing principle, the \textit{open-source hardware} model offers a promising avenue to streamline and accelerate the development of new SoCs, both in terms of cost and time. The principle, exemplified in Fig.~\ref{fig:open_model}, is simple: instead of allocating significant resources to integrate outsourced IPs from vendors for low-value common baseline, non-differentiating parts of an SoC, one can focus efforts and funding primarily on the development of differentiating proprietary IPs and outsource only those technology-dependent IPs that are of critical importance (e.g., DRAM PHYs).
Moreover, one can leverage available high-quality open-source IPs as a ``starting point'' for their designs, avoiding the need to fund development from scratch.

Since 2013, the academic PULP (\textit{Parallel Ultra-Low Power}) Platform project has been one of the most active and successful initiatives in designing research IPs and releasing them as open-source. Its portfolio now ranges from processor cores to network-on-chips, peripherals, SoC templates, and full hardware accelerators\footnote{https://github.com/pulp-platform}.
In this article, we focus on the PULP experience designing heterogeneous AI acceleration SoCs -- an endeavour encompassing SoC architecture definition; development, verification, and integration of acceleration IPs; front- and back-end VLSI design; testing; development of AI deployment software.

\section*{Heterogeneous PULP SoCs}
\label{sec:heterogeneous_pulp}

\begin{figure*}[tb]
    \centering
    \includegraphics[width=0.75\linewidth]{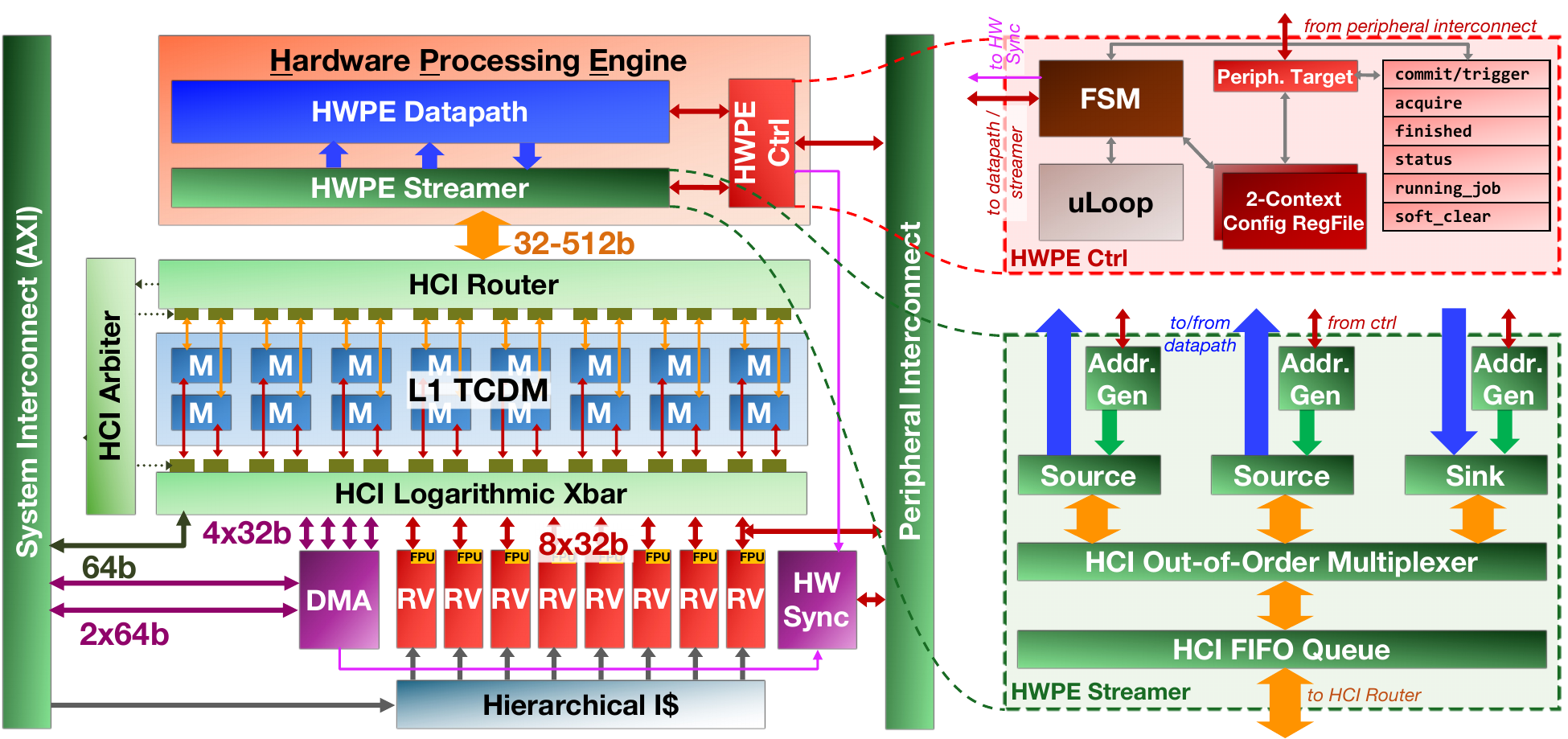}
    \caption{
    \textbf{Left}: template of a heterogeneous PULP cluster. \textbf{Right}: internal organization of an HWPE \textit{streamer} and \textit{controller} blocks.
    }
    \label{fig:heterogeneous_cluster}
\end{figure*}

The seed idea behind the Parallel Ultra-Low-Power (PULP) Platform architecture is to target a low-voltage, low-frequency, but highly energy-efficient operating point~\cite{deNearthresholdVoltageDesign2013} and compensate for performance loss through architectural parallelism and hardware acceleration rather than relying on high-frequency operation.
Therefore, the architecture of the main PULP computing block -- called a PULP \textit{cluster} -- is designed to leverage cooperation between programmable parallel processors (potentially with ISA extensions) and fixed-function HW accelerators that aim at the maximization of efficiency for a performance target.

The heterogeneous PULP cluster template, shown in Fig.~\ref{fig:heterogeneous_cluster}, is organized around three main fundamental blocks.
The first block, central to the operation of the cluster, is a multi-banked L1 Tightly-Coupled Data Memory (TCDM), typically in the range 64--256KiB of size with 16--64 banks, each with 32-bit wide ports.
The second block is a set of RISC-V processors (4--16); each core has its own 32-bit load/store data memory port, addressing one bank at a time.
RISC-V cores may employ Instruction Set Architecture (ISA) extensions for digital signal processing and AI~\cite{contiMarsellusHeterogeneousRISCV2024,nadaliniTOPSRISCVParallel2023}.
The third block is a subsystem with one or more Hardware Processing Engines (HWPEs), time-sharing a single wide memory port ($N\times$32-bit) targeting many memory banks with the same access.
The ``connective tissue'' of the cluster is a Heterogeneous Cluster Interconnect (HCI) that arbitrates between HWPE and core-side accesses and routes HWPE accesses to the correct set of banks~\cite{prasadSpecializationMeetsFlexibility2023}.
The HCI is organized into two branches: a logarithmic crossbar to enable fair access from each one of the RISC-V cores and DMA to 32b banks; and a router for wide (up to 512b) access from HWPEs to multiple banks at a time.
The cluster is completed by a set of peripherals, the two foremost of which are a DMA engine and a hardware synchronizer that is used to enable synchronization primitives, such as fast barriers and critical sections.

PULP clusters are generally integrated as mixed hardware/software acceleration engines inside larger SoCs: in many cases, relatively simple microcontrollers; in others, more complete SoCs designed for integration in single-board embedded computers.
Different flavours of PULP clusters (with different HWPEs or ISA extensions, memories, etc.) can be integrated within the same SoC to accelerate a diversity of tasks, creating a ``fractally heterogeneous'' architecture.

\subsection*{ISA Extensions for AI}

\begin{figure}[tb]
    \centering
    \includegraphics[width=0.95\linewidth]{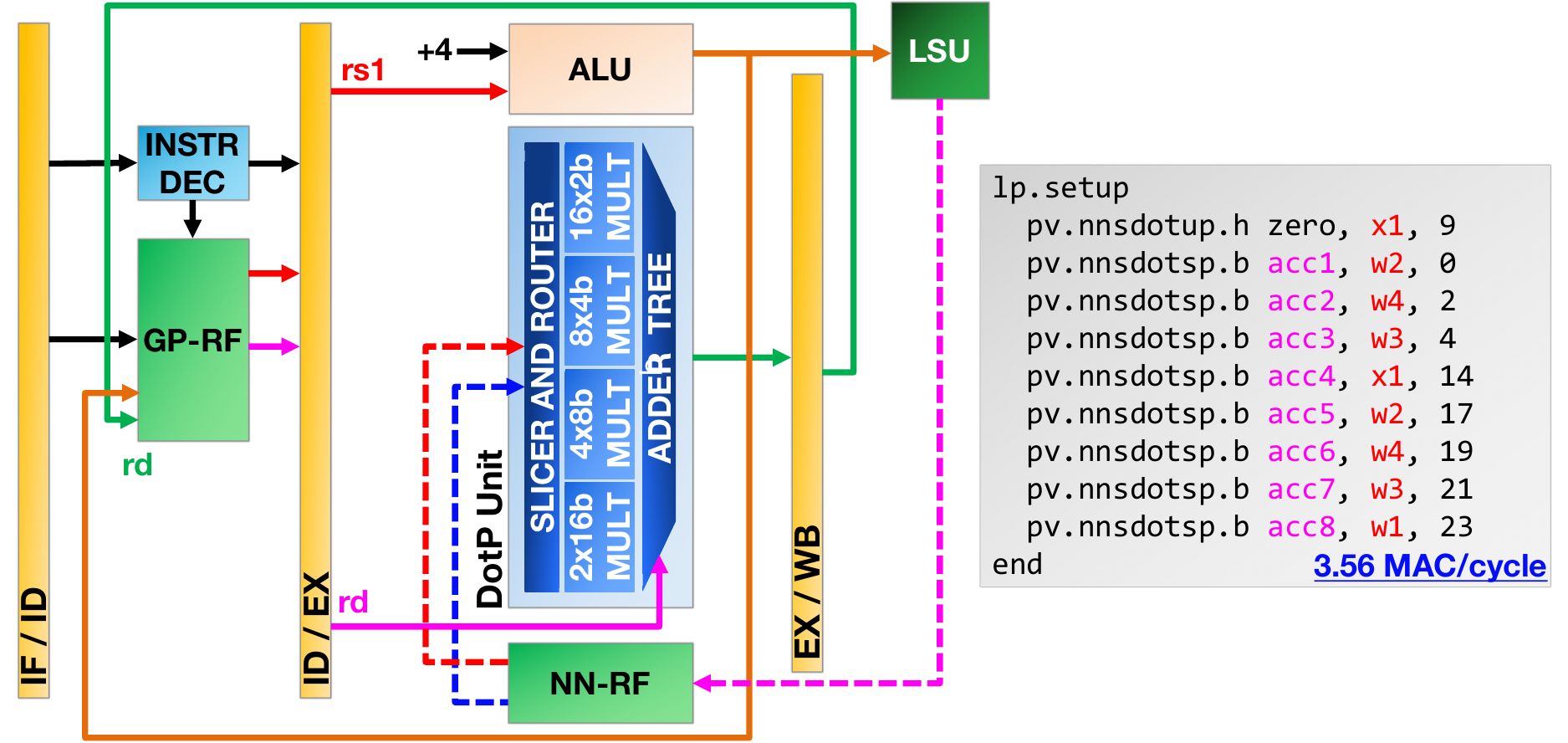}
    \caption{\texttt{Xpulpnn} datapath integrated in RI5CY in the \textit{Marsellus} SoC. The special-purpose \textit{NN-RF} register file is fed by the load-store unit and used to feed the dot-product unit without impacting the general purpose register file (\textit{GP-RF}), which enables to considerably improve internal data reuse.}
    \label{fig:xpulpnn_marsellus}
\end{figure}

As previously mentioned, the first heterogeneity strategy employed in PULP clusters is to improve the architectural efficiency of the cluster by selecting key operations, such as dot-products, that can be implemented as single instructions.
ISA extensions significantly boost performance \& energy efficiency while maintaining the flexibility of the baseline instruction  processors.
The modularity of the RISC-V ISA, which we use in PULP clusters, allows for the seamless addition of custom instructions tailored to specific computational tasks.
All PULP-based chips using the RI5CY/CV32E40P core employ the \texttt{Xpulp} ISA extension~\cite{gautschiNearThresholdRISCVCore2017}, which includes a set of instructions for digital signal processing (DSP), such as 8-bit SIMD dot-products, 32-bit multiply-accumulate, hardware loops, load/store with post-increment.

A more advanced set of instruction set extensions are those that target the optimization of Quantized Neural Network (QNN) workloads, where low-bit-width arithmetic (e.g., 2-bit to 8-bit operations) is commonly employed to reduce memory and computational demands.
Over the years, we developed sophisticated techniques such as low-bitwidth dot-products (\texttt{Xpulpnn}~\cite{garofaloXpulpNNEnablingEnergy2021,contiMarsellusHeterogeneousRISCV2024}), status-based instructions with automatic MAC\&Load~\cite{nadaliniTOPSRISCVParallel2023}, and lockstep execution~\cite{ottaviDustin16CoresParallel2023}.
Fig.~\ref{fig:xpulpnn_marsellus} shows the datapath of \texttt{Xpulpnn} integrated inside the RISC-V cores in the \textit{Marsellus} SoC, which includes a special-purpose register file with a dedicated connection to the LSU and a multi-precision dot-product unit, enabling ultra-tight execution loops with no overheads for load/store, looping, etc.
Support for these enhanced instructions can be added to existing RISC-V toolchains, such as LLVM and GCC.

\begin{figure*}[tb]
    \centering
    \includegraphics[width=0.99\linewidth]{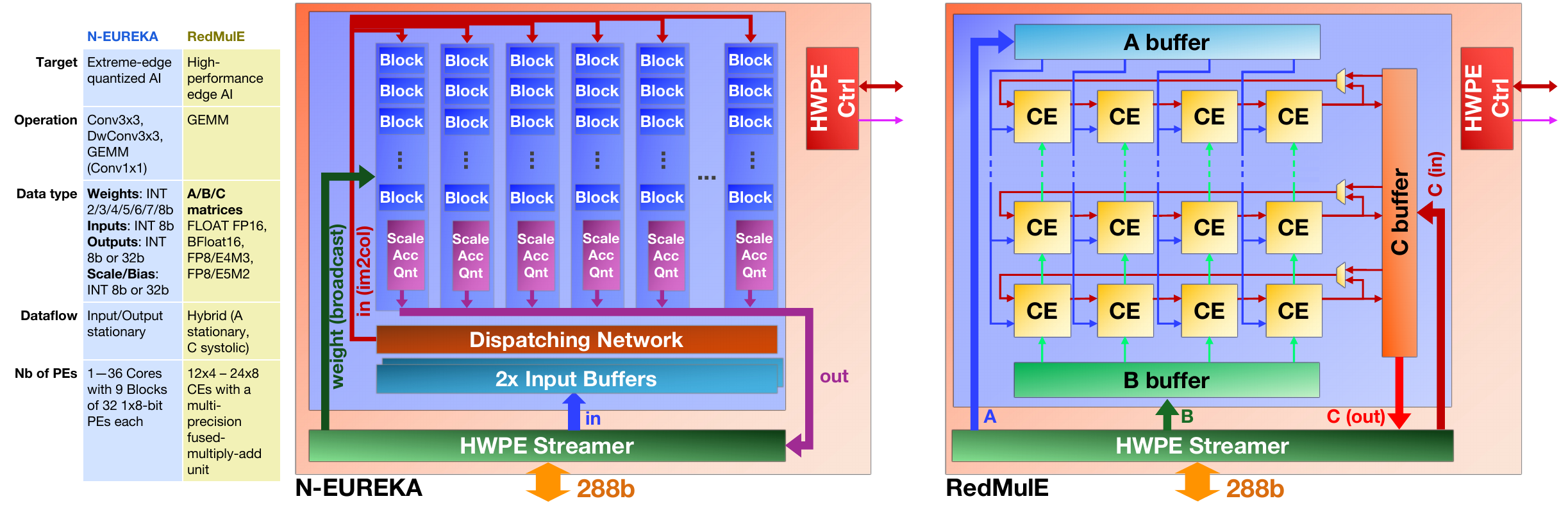}
    \caption{Example of two HWPE architectures. \textbf{Left}: N-EUREKA, dedicated to extreme-edge quantized AI; \textbf{right}: RedMulE, dedicated to high-performance edge inference \& training.}
    \label{fig:hwpe_examples}
\end{figure*}

When ISA extensions overshoot a certain level of complexity (e.g., instruction execution takes a number of cycles significantly $>$1 and/or requires significantly larger memory bandwidth than the peak 32b/cycle of a simple RV32 core), we prefer to exploit a different heterogeneity strategy with cluster-coupled cooperative HWPEs.
In this way, RISC-V cores are kept relatively lean, flexible, and energy-efficient. At the same time, more aggressive specialization in HWPEs is not bound to the strict fetch-decode-execute approach of a processor.

\subsection*{Hardware Processing Engines}
The idea behind the HWPE approach is simple and based on the Pareto principle: accelerate $\sim$20\% of the total workload that takes $\sim$80\% of the execution time.
In AI applications, these workloads include primarily convolutional layers in CNNs and matrix multiplications in other DNNs and Transformers.
Whereas in most academic and industrial examples of edge AI large accelerators are coupled to a small general-purpose processor, HWPEs are meant to be used cooperatively with the cluster of RISC-V cores.
The cores are less specialized than the HWPEs, but still capable of considerable acceleration thanks to parallel computing and ISA extensions.
The fact that both ``critical'' and less critical tasks are accelerated, albeit to different degrees, results in a balanced system.

HWPEs\footnote{https://hwpe-doc.rtfd.io} are structured in three standard blocks, to enhance reusability in an open-source hardware setting: \textit{controller}, \textit{streamer}, and \textit{engine} or \textit{datapath}, as highlighted in Fig.~\ref{fig:heterogeneous_cluster}.
All controllers combine custom logic with open-source modules\footnote{https://github.com/pulp-platform/hwpe-ctrl} to implement common functionality.
The \textbf{controller} enables cores to configure the HWPE, via a memory-mapped target port; it also controls all other accelerator components.
The offload of a job from a core to the accelerator consists of \textit{acquiring} a lock, writing parameters to the internal HWPE register file, and then \textit{triggering} the execution.
The controller also supports auxiliary functions such as soft-clearing the state of the HWPE, checking its current status, and getting the ID of the currently running job. 
Moreover, the register file supports multiple contexts to overlap the programming of a new job with the execution of the previous one.
The \textbf{streamer} is a custom DMA engine specialized in enqueuing data from memory into ready/valid latency-tolerant internal \textit{streams} towards the HWPE datapath; and in dequeuing into towards memory data from streams produced by the datapath.
Streamers are generally composed entirely out of a collection of open-source modules such as FIFO queues, multiplexers, address generators, and stream sources/sinks that are designed to plug into the HCI interconnect\footnote{https://github.com/pulp-platform/hci}.
Streamers are designed to smooth the interface between accelerators and the cluster TCDM while providing high-bandwidth access (up to 512b/cycle in current-generation HWPEs).

\begin{figure*}[tb]
    \centering
    \includegraphics[width=0.85\linewidth]{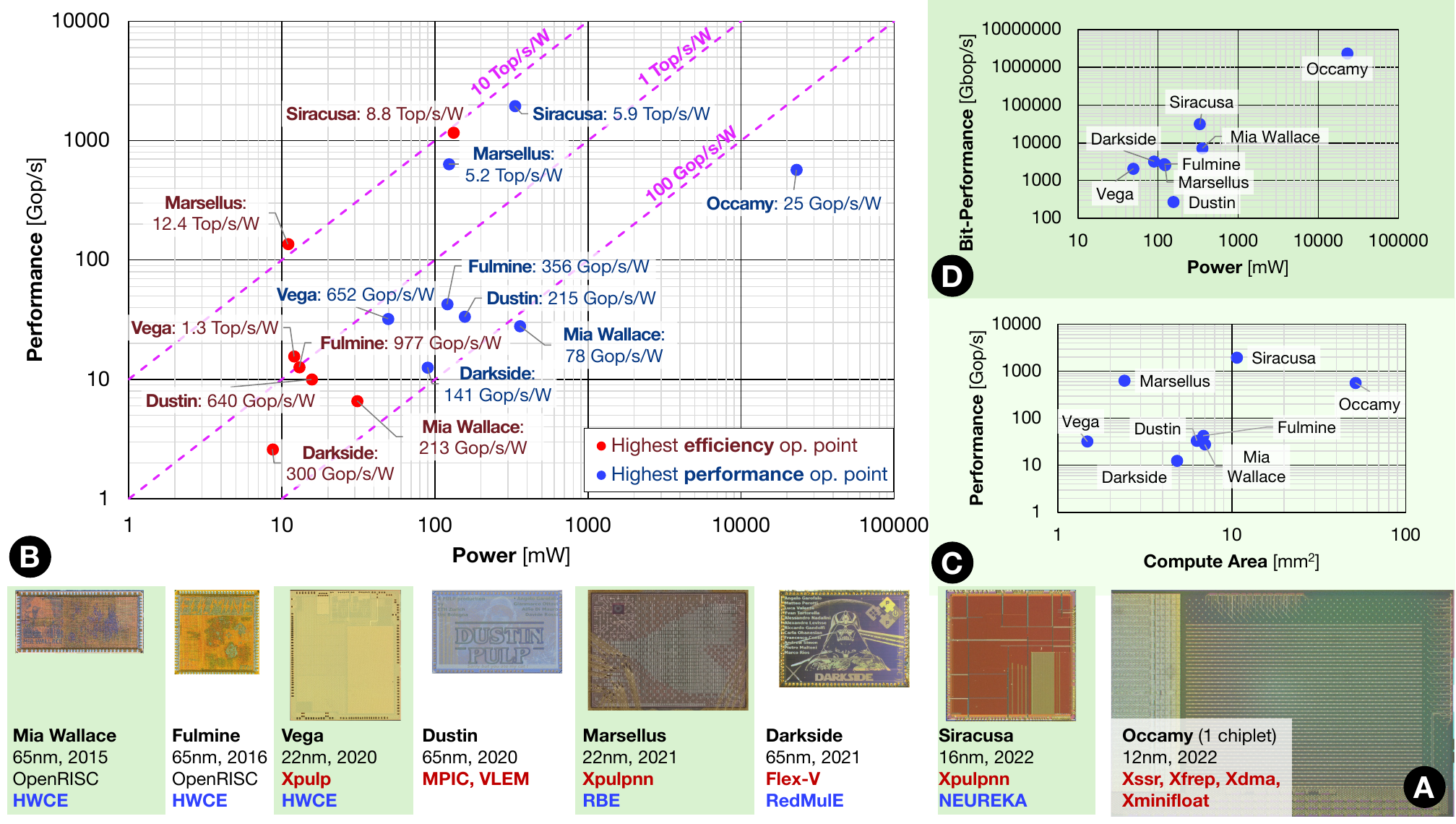}
    \caption{Analysis of a selection of PULP chips taped out between 2015 and 2022, exploiting architectural heterogeneity as ISA extensions (\textbf{\color{red}red}) or HWPEs (\textbf{\color{blue}blue}).
    \textit{\textbf{A)}} Microphotographs of the taped-out prototypes (in-scale with one another); \textit{\textbf{B)}} Performance (Gop/s), power (mW), and energy efficiency of the considered PULP SoCs in their respective highest performance and energy efficiency operating points; \textit{\textbf{C)}} Peak performance (Gop/s) versus SoC area devoted to computation;  \textit{\textbf{D)}} Normalized performance per bit (e.g., 1 Gop/s @ 4b-weight, 8b-input precision = 32 Gbop/s) versus power (mW) in the highest performance operating point.}
    \label{fig:pulp_chips}
\end{figure*}

The \textbf{datapath} of accelerators, contrary to streamer \& controller, is always custom-tailored for a given application.
In Fig.~\ref{fig:hwpe_examples}, we detail the datapath of two entirely open-source HWPEs designed for AI acceleration: respectively, \textbf{N-EUREKA} for extreme edge quantized AI and \textbf{RedMulE} for GEMM in high-performance edge AI with floating-point data.
The two accelerators target  different use cases with diverse datapaths. Still, they are connected to the rest of the PULP heterogeneous cluster with the same strategy, employing HWPE streamers and controllers.
\textbf{RedMulE}~\cite{tortorellaRedMuleMixedprecisionMatrix2023}, shown right, is a systolic $M\times N$ array dataflow accelerator for FP16/BFloat16/E4M3/E5M2 generalized matrix multiplication ($\mathbf{C} \pluseq \mathbf{A}\cdot \mathbf{B}$). $\mathbf{A}$ matrix elements are stationary and dispatched to $M\times N$ \textit{computing elements} (CEs) according to their address: each CE uses a different  stationary $A_{i,j}$ element. $\mathbf{B}$ matrix elements are streamed in continuously and broadcast to the $M$ rows of CEs, e.g., all CEs in the $j$-th column will use the same $B_{t,j}$ in cycle $t$.
Finally, $\mathbf{C}$ matrix elements are circulated systolically throughout each of the $M$ rows: after a CE in column $j$ has computed its contribution to $C$, it passes it to the CE on its right ($j+1$), or to the first CE at the end of the array.
To enable pipelining of the fused-multiply-accumulate units used inside CEs, CEs operate with a latency of 4 cycles and a throughput of 1 operation per cycle.
In \textit{Darkside}~\cite{garofaloDARKSIDEHeterogeneousRISCV2022}, we included a $12\times 4$-CE version of this HWPE.

The HWPE approach towards integration into a PULP system is generic and not tied to a specific accelerator datapath.
In fact, even accelerator datapaths that are entirely different from regular digital designs, such as analog in-memory computing (AIMC) arrays, can be integrated as HWPEs.
In Garofalo~et~al.~\cite{garofaloHeterogeneousInMemoryComputing2022}, we explored the case of an AIMC accelerator based on phase-change memory (PCM): as long as the inputs and outputs of the datapath can be digitized and encoded in input and output streams, it can be encapsulated inside an HWPE.
The advantage is that 30--60\% of the code can be reused between different HWPE designs.
Moreover, also the hardware abstraction layer (HAL) that programs the HWPEs can be largely reused between different designs, enabling a fast cycle from new microarchitecture ideas to deployment in complete PULP systems.

\subsection*{Heterogeneous PULP SoCs across the ages}
We started the development of the PULP platform architecture in 2013, and the earliest SoC including architectural heterogeneity (already targeted towards AI, and specifically CNNs) is from 2015.
We have come a long way since then!
In Fig.~\ref{fig:pulp_chips}, we show a selection of heterogeneous PULP chips for AI across time, with their evolution in architectural features, performance, and energy efficiency.
Older SoCs (\textit{Mia~Wallace}~\cite{pulliniHeterogeneousMultiCoreSystemonChip2017}, \textit{Fulmine}~\cite{contiIoTEndpointSystemonChip2017}) exploit iterations of the RI5CY core not yet relying on the RISC-V ISA and include an early embodiment of the HWPE concept for convolutional layer acceleration (HWCE).
A newer iteration of the same design was also exploited in \textit{Vega}~\cite{rossiVegaTenCoreSoC2022}, which included also the first \texttt{Xpulp} DSP extensions.

\textit{Marsellus}~\cite{contiMarsellusHeterogeneousRISCV2024} and \textit{Siracusa}~\cite{prasadSiracusa16Nm2024} introduced newer generation HWPEs (RBE, N-EUREKA) with similar architectures oriented towards aggressively quantized inference.
Both chips also employed the \textit{Xpulpnn} ISA extension presented in Fig.~\ref{fig:xpulpnn_marsellus}, to cover layers that are not efficiently accelerated by the HWPEs.
\textit{Siracusa} also includes what we call \textit{at-memory computing}, with the N-EUREKA HWPE tightly coupled with a non-volatile on-chip MRAM used for DNN weights.
These two SoCs constitute the current ``PULP'' state-of-the-art for what concerns hardware-accelerated AI.

\textit{Dustin}~\cite{ottaviDustin16CoresParallel2023} does not include HWPEs, but it is one of the most aggressive low-power PULP SoCs in terms of ISA extensions, including multi-precision integer convolution (MPIC), which extends the \texttt{Xpulpnn} concept to include asymmetric precision dot-product (e.g., 4$\times$8b) and features GPU-style vector lockstep execution mode (VLEM) that dramatically reduces instruction fetch/decode overheads.
\textit{Darkside}~\cite{garofaloDARKSIDEHeterogeneousRISCV2022} is the first SoC using the previously discussed RedMulE HWPE, and it also features the Flex-V cores~\cite{nadaliniTOPSRISCVParallel2023}, which combine the \texttt{Xpulpnn} and MPIC concepts in a single design.

All of these designs are extreme edge AI designs: for comparison, and to indicate the road ahead for the PULP SoCs, Fig.~\ref{fig:pulp_chips} also reports results for \textit{Occamy}~\cite{paulinOccamy432Core2812024}.
Occamy is a dual-chiplet, many-core design featuring a total of 432 Snitch cores (organized in 48 clusters) with aggressive ISA extensions for low-precision floating point and sparse stream semantic registers (SSRs).
SSRs work similarly to HWPE streamers, but direct their access streams towards functional units tightly-coupled to the Snitch cores: this introduces a middle-ground between conventional ISA extensions that exploit regular processor resources (instruction fetch/decode units and load/store units) and HWPEs that have no instructions and fully customize their load/store units.
Targeted at a high-performance computing scenario with FP64 support rather than at pure AI acceleration, Occamy does not aim at being directly competitive in AI inference with aggressively quantized designs such as Siracusa and Marsellus in terms of energy efficiency in the $\sim$10TOPS/W range; and shows a similar efficiency (in the $\sim$100GOPS/W range) to Darkside, a hardware-accelerated SoC in a much older technology node (12nm vs 65nm).
Nevertheless, its design opens a pathway for the future scalability of fully open-source hardware-accelerated PULP SoCs, as we discuss at the end of this article.

\subsection*{PULP in the AI State-of-the-Art}

\begin{figure}[tb]
    \centering
    \includegraphics[width=0.95\linewidth]{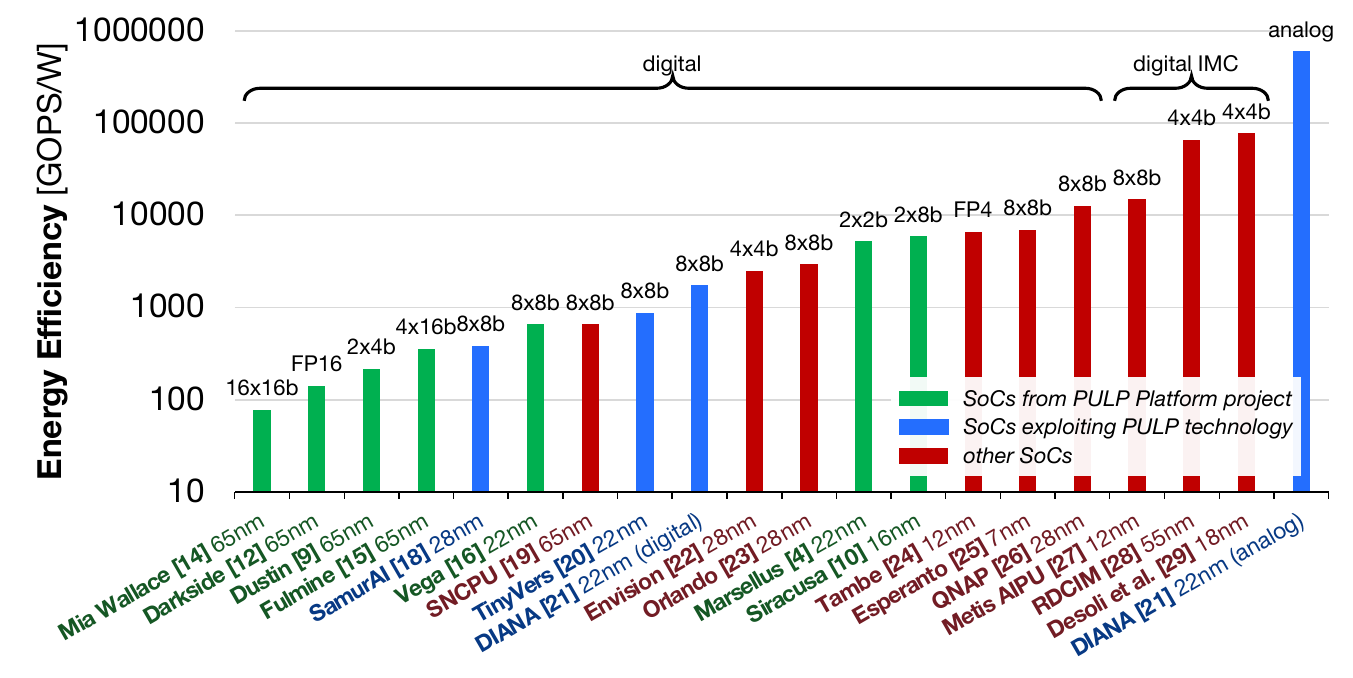}
    \caption{Efficiency of AI Systems-on-Chip measured in peak-performance operating point.}
    \label{fig:soa}
\end{figure}

\nocite{pulliniHeterogeneousMultiCoreSystemonChip2017,garofaloDARKSIDEHeterogeneousRISCV2022,ottaviDustin16CoresParallel2023,contiIoTEndpointSystemonChip2017,miro-panadesSamurAIVersatileIoT2022,rossiVegaTenCoreSoC2022,juSystolicNeuralCPU2023,jainTinyVersTinyVersatile2023,houshmandDIANAEndtoEndHybrid2023,moonsEnvision26to10TOPSSubwordparallel2017,desoli9TOPSDeepConvolutional2017,contiMarsellusHeterogeneousRISCV2024,prasadSiracusa16Nm2024,tambe2212nm182023,ditzelAcceleratingMLRecommendation2022,mo12TOPSQuantized2022,hager113MetisAIPU2024,yiRDCIMRISCVSupported2024,desoli16740310TOPSSRAMBased2023}

One of the big advantages of open-source hardware is that it can be an ``innovation booster'' for many more people than conventional closed approaches. However, to really push this argument, it is essential to show that solutions based on open-source hardware are competitive with closed ones.
In Fig.~\ref{fig:soa}, we compare (in terms of energy efficiency) the previously discussed PULP SoCs stand within the state-of-the-art of AI-oriented heterogeneous SoCs.

Not only the most advanced PULP heterogeneous SoCs (\textit{Marsellus}, \textit{Siracusa}) are competitive with the peak of the state-of-the-art for purely digital designs; but several of the non-PULP SoCs actually exploit PULP technology inside (\textit{SamurAI}~\cite{miro-panadesSamurAIVersatileIoT2022}, \textit{TinyVers}~\cite{jainTinyVersTinyVersatile2023}, \textit{Diana}~\cite{houshmandDIANAEndtoEndHybrid2023}).
Employing analog in-memory computing combined with digital acceleration and PULP technology, \textit{Diana} shows the highest energy efficiency (up to $\sim$600 TOPS/W).

\section*{AI Deployment flows}

The architecture and SoC design is only one half of the design issue for complex SoCs, and open-source ones make no exception.
Heterogeneous PULP SoCs are designed to exploit the presence of many programmable cores.
%
This means that most of the complexity of control of HWPEs, DMAs, and other blocks is delegated to the program running on the RISC-V cores in the system.
Flexibility is maximum, but there's a problem: who writes this code?

\subsection*{Maximizing hardware utilization with tiling}
We split this hard problem into two separate smaller challenges.
The first is to write small dedicated \textit{back-end kernels} that run directly on the cluster.
These are typically (relatively) simple, dedicated to  a single task (e.g., a convolutional layer), and assume that all data is available in the nearest local memory, i.e., the TCDM.
Kernels are hand-coded to make optimal use of available resources, without explicit care for data transfers to other levels of the memory hierarchy.
To make sure that all data is available locally to a kernel, we split workloads deployed to a cluster in \textit{tiles}, i.e., units of work (and data) that can be executed independently from one another and sized appropriately to fit within the available on-cluster resources.
AI workloads, dominated by linear algebra, can be tiled well due to their regularity.

\begin{figure}[tb]
    \centering
    \includegraphics[width=0.95\linewidth]{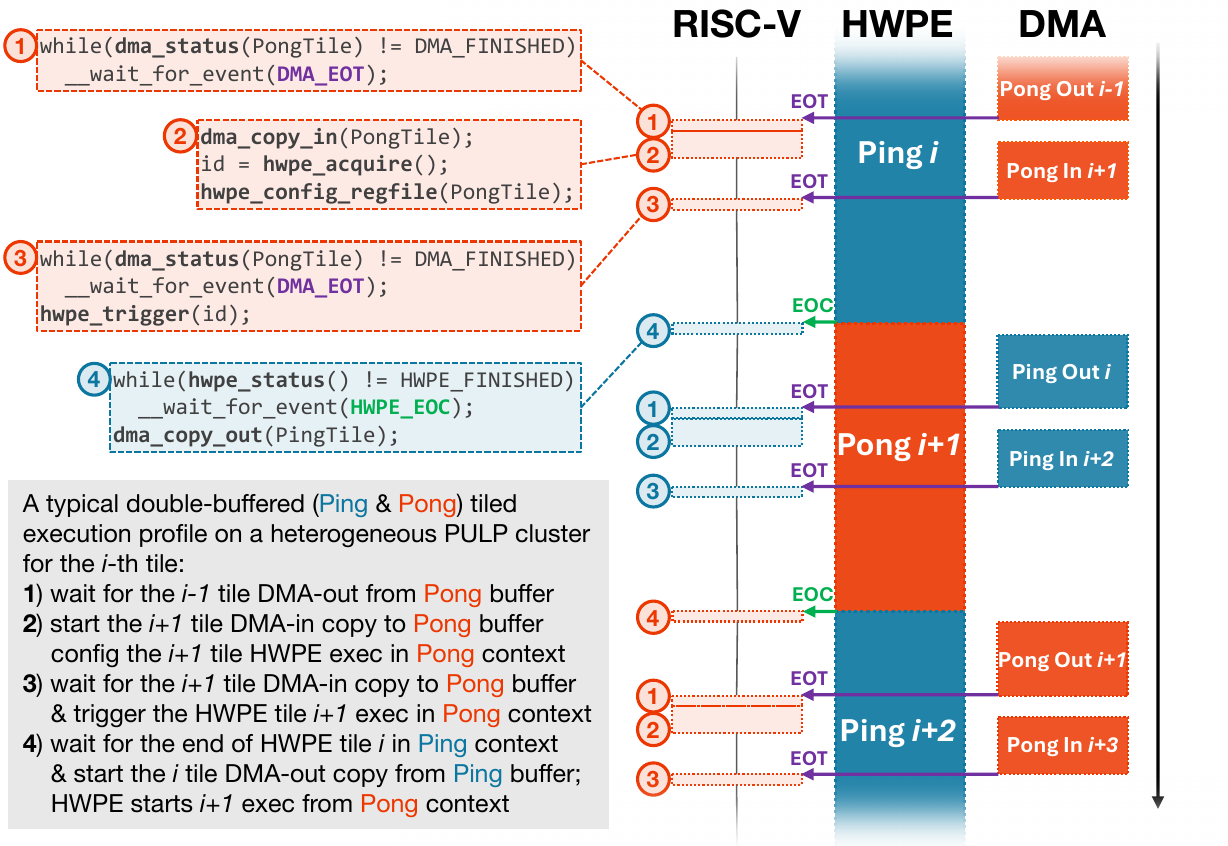}
    \caption{Tiled execution profile for a HWPE-based computation.}
    \label{fig:hwpe_db_model}
\end{figure}

\begin{figure*}[tb]
    \centering
    \includegraphics[width=0.75\linewidth]{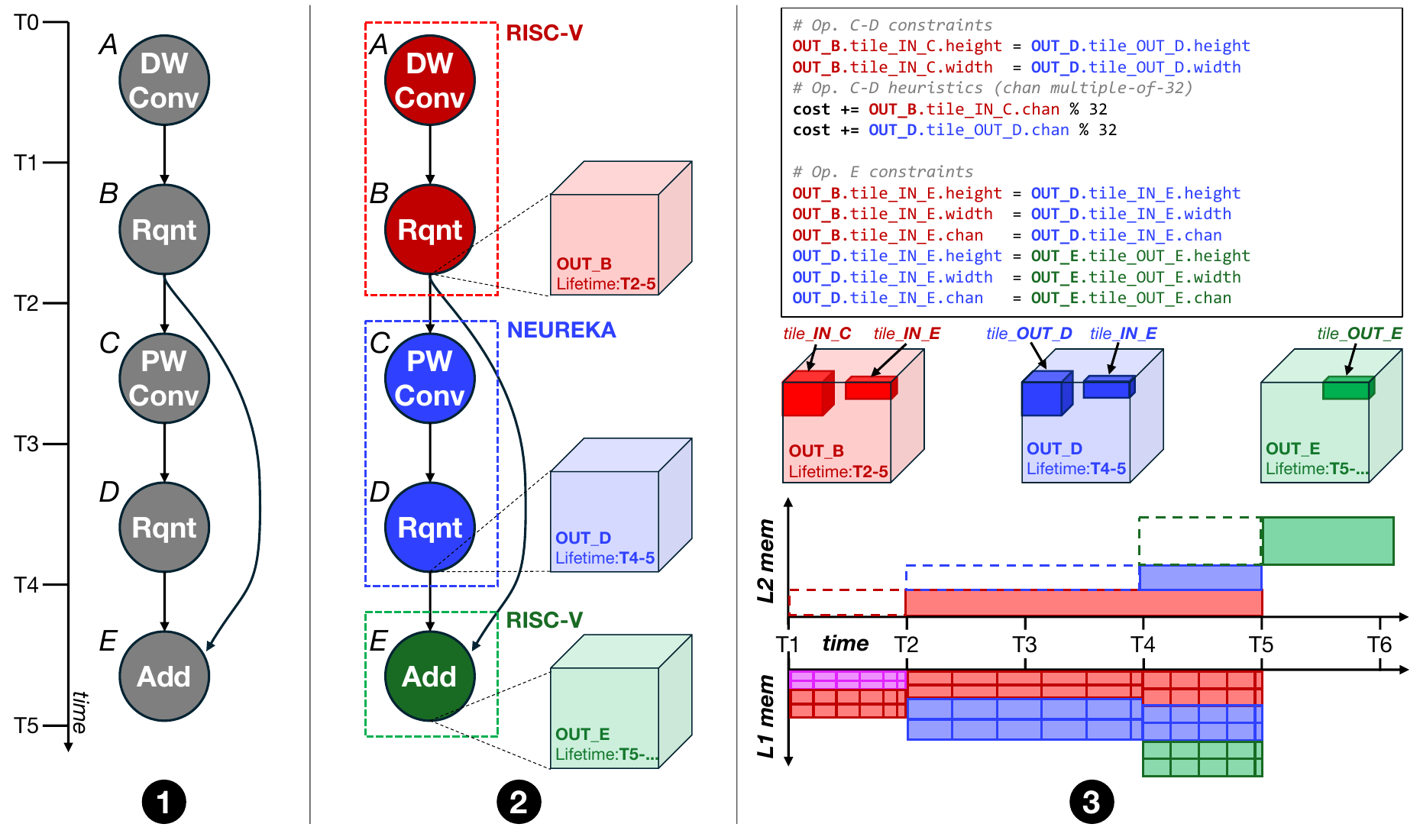}
    \caption{Three stages of the tiling \& deployment procedure in the example of Siracusa: \textit{\textbf{1)}} The quantized network, represented as an ONNX graph, is mapped to an internal representation (nodes A--E). \textit{\textbf{2)}} Nodes are \textit{fused} and \textit{colored} according to the target operator; input and output tensors are tagged with their lifetime and promoted to their target memory. \textit{\textbf{3)}} Tensor lifetimes and geometrical constraints are combined to schedule \& allocate tensors and time buffers in all system scratchpads (L1, L2, etc.).}
    \label{fig:deployment}
\end{figure*}

A hand-written kernel will eventually get executed in a loop that uses tiling and double-buffering to hide transfer overheads and maximize hardware utilization.
As an example, Fig.~\ref{fig:hwpe_db_model} shows the case of a layer that executes mainly a kernel on a HWPE, but with essential assistance from the cluster DMA and one of the RISC-V cores to implement a fully double-buffered scheme designed to keep the HWPE always busy.
The underlying hardware mechanisms are all provided directly by the HWPE controller, the DMA, and the cluster's hardware synchronizer.
A typical iteration $i$ of a tiled execution loop waits for the previous tile's ($i-1$) DMA copy-out, starts the next one's ($i+1$) copy-in, and programs the HWPE for $i+1$.
The $i+1$ HWPE job can be triggered as soon as the copy-in operation is finished.
Finally, after the current tile $i$ execution has also finished, the next one automatically starts from the HWPE controller job queue.
The RISC-V core is notified of end-of-transfer (EOT) and end-of-computation (EOC) events via the cluster synchronizer.
%
More complex schemes can also be implemented; e.g., it is possible for HWPEs and RISC-V cores to work simultaneously sharing access to the TCDM through the HCI interconnect.

\subsection*{Generating the tiling code}
This approach leaves out two questions: who decides the tiling grid? And how is the code implementing the tiled execution profile of Fig.~\ref{fig:hwpe_db_model} written?
We may accept hand-coding reusable back-end kernels, but we certainly do not want to manually code all the tiling loops.
Most open-source deployment flows such as TVM~\cite{chenTVMAutomatedEndtoEnd2018} focus on end-to-end top-down code optimizations that implicitly assume a regular data cache hierarchy, with a ``flat'' memory view.
Instead, already in current-generation PULP heterogeneous SoCs, we have complex hierarchies of manually managed scratchpads (and future chips will employ more complex ones).
We developed two generations of custom tools that are tailored to the needs of our architectures: DORY~\cite{burrelloDORYAutomaticEndtoEnd2021}, targeting homogeneous PULP clusters (without HWPEs), and Deeploy~\cite{schererDeeployEnablingEnergyEfficient2024}, which supports the integration of HWPEs and scales to more complex SoC architectures (e.g., multiple clusters).

Fig.~\ref{fig:deployment} details the strategy that our AI deployment flows exploit.
The main goal of these tools is to transform a high-level representation of the target work into low-level C code that includes all kernel function calls, data movement, HWPE programming, and synchronization points -- all while maximizing the utilization of hardware units (cores and HWPEs).
We hand-tune computational software kernels for the RISC-V cluster and for HWPEs so that we can  exploit more effectively ISA extensions without requiring compiler-level support for advanced optimizations based on the ISA semantics.
These kernels are relatively straightforward to write because they assume all data is allocated in L1, and they do not require explicit data movement.

\begin{figure}[tb]
    \centering
    \includegraphics[width=0.95\linewidth]{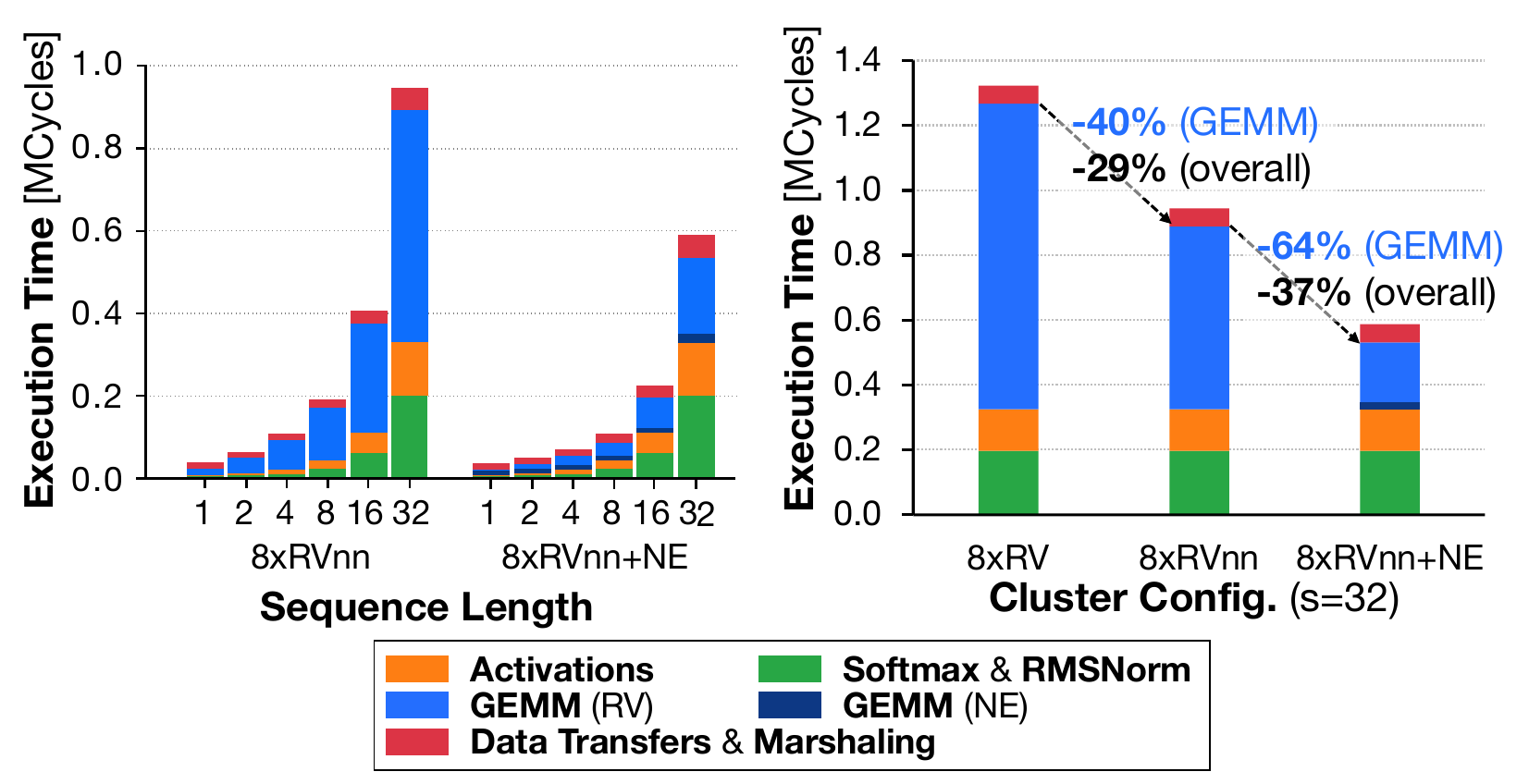}
    \caption{Example of execution of an end-to-end encoder Transformer~\cite{schererDeeployEnablingEnergyEfficient2024} in a PULP cluster with three configurations: \texttt{Xpulp} DSP ISA extensions (\textit{8xRV}); \texttt{Xpulpnn} AI ISA extensions (\textit{8xRVnn}); and with both AI ISA extensions and N-EUREKA (\textit{8xRVnn+NE}). The Transformer has 8 layers, hidden size $d_m=64$, $h=16$ parallel heads, intermediate size for feed-forward layers $d_{ff}=256$. Left: execution time with sequence length sweeping from $s=1$ to $s=32$; right: detail of $s=32$ showing the speedup from pure \texttt{Xpulp}-based execution to \texttt{Xpulpnn} and N-EUREKA.}
    \label{fig:deeploy_encoder}
\end{figure}

Knowing a target back-end library of kernels, the deployment tool takes a computational graph of an application and maps it to an internal representation so that each node can be optimized (by fusing or splitting them) and ``colored'' to be allocated to one of the available computing engines.
The inputs and outputs are then allocated to buffers in various levels of the memory hierarchy in which they can be entirely materialized, deriving a schedule of buffer lifetimes.
At the end of this step, a tiling grid and schedule need to be defined, which in turn will be used to define all synchronization requirements and data transfers that are needed.
Our tools set up the search for near-optimal tiling grids as a Constraint Programming (CP) problem, putting together geometric constraints from the AI model layers, known tensor lifetimes, and heuristic cost factors modeling tile size preferences on the back-end side (e.g., tile sizes aligned to the microarchitecture of the target HWPE).
The solution to this CP problem is a schedule \& allocation of tiling buffers across all the memories in the SoC, considering also double buffering at all memory levels.
This is used by a final code generation stage, which generates a C source file putting together data movement, synchronization, and calls to the back-end kernels, which can then be compiled using a standard C compiler for RISC-V (e.g., LLVM) extended to support our ISA extensions.

Fig.~\ref{fig:deeploy_encoder} shows the example of the results of the result of this process in terms of end-to-end execution of a Transformer (encoder) AI model on \textit{Siracusa}, as reported in Scherer~et~al.~\cite{schererDeeployEnablingEnergyEfficient2024}, utilizing the Deeploy deployment tool.
This is a particularly informative example because of the complexity of the task, which includes a variety of  layers: projection GEMMs can be deployed on N-EUREKA or exploit \texttt{Xpulpnn} and parallel acceleration in software; attention GEMMs can only exploit \texttt{Xpulpnn} and parallelism; and other layers, such as softmax, can only employ parallel acceleration.
Even when using all acceleration techniques, the overhead for data transfers \& marshaling (which includes the cost of all the code generated by Deeploy) is kept $<$10\% on the full network execution, which means that the goal of maximal hardware utilization is achieved.

All compilers, software development kits (SDKs), and deployment tools developed in the PULP platform project are entirely open-source, similar to what we do for hardware.
Software design costs are often reported as equal to or higher than hardware design ones for SoCs in advanced technology nodes~\cite{EdgeAIVision2023}: access to a mature hardware/software design template can make the difference for small enterprises and startups wishing to design an SoC from scratch.

\section*{The future: scaling-up heterogeneity (and AI)}
What we discussed so far is the present of PULP for AI. But what about the future? As discussed, the computational needs of AI scale faster than our capability to design systems to run AI on.
We believe open-source SoCs are essential to foster more innovation on AI acceleration and make AI more accessible, private, and secure with on-device local execution -- and that PULP has a role to play in that game by enabling the design and prototyping of academic/open-source SoCs of unprecedented complexity, and releasing the results to academia and industry all over the world.

\textit{Occamy}~\cite{paulinOccamy432Core2812024} shows the road ahead in this regard, with a design that is many times more complex than the other PULP-based SoCs.
It is the first open-source design demonstrating performance scalability on a large-scale many-core  system, integrating high-performance node-locked IPs (high-bandwidth memory), and exploiting System-in-Package integration.
One of the main elements of our vision for next-generation heterogeneous PULP systems is to combine this intuition with the advantages of heterogeneous acceleration that we demonstrated in smaller chips, creating next-generation fully open-source high-performance edge AI SoCs.
While so far higher-performance platforms have been mainly the domain of proprietary architectures, open designs can be key to democratizing AI hardware as they did for many other aspects of computing, making local inference and training possible at a fraction of the cost in the future.

An orthogonal element that is conspicuously not yet fully exploited in PULP SoCs is the integration of advanced technology IPs, such as digital or analog in-memory computing (IMC) arrays.
Numerous works in the state-of-the-art have demonstrated that tighter in-memory computing approaches can deliver better energy efficiency, albeit typically at the cost of more aggressive customization.
The heterogeneous approaches discussed here can be applied to integrate complex IMC arrays just as digital datapaths.
``At-memory computing'' applied in Siracusa with MRAM can be thought of as only one step removed from the integration of digital blocks inside a non-volatile memory; more aggressive pursuits could lead to improved computational energy efficiency and, especially in the case of NVM, to relieve the memory-boundedness of some of the emerging generative AI applications, such as inference of decoder-only Transformers.

The third, big avenue of evolution for open-source AI designs, and specifically our designs based on PULP, are open-source process design kits (PDKs) and electronic design automation (EDA) tools.
Bringing down the cost of development of new SoCs by an order of magnitude, the availability of open-source tools and (relatively) cheap prototyping might be a paradigmatic change, inducing the development of a much more varied and heterogeneous ecosystem for AI, with tailored accelerators for each specific need.


\section*{Acknowledgment}
The authors would like to thank the entirety of the PULP team at ETH Z\"urich, the University of Bologna,  and ``at large'', as well as the Design Zentrum of ETH Z\"urich, and in particular Frank K. G\"urkaynak, for their support in designing and taping out all PULP SoCs (and taking pictures!).

\ifCLASSOPTIONcaptionsoff
  \newpage
\fi



\bibliographystyle{IEEEtran}

\begin{thebibliography}{10}
\providecommand{\url}[1]{#1}
\csname url@samestyle\endcsname
\providecommand{\newblock}{\relax}
\providecommand{\bibinfo}[2]{#2}
\providecommand{\BIBentrySTDinterwordspacing}{\spaceskip=0pt\relax}
\providecommand{\BIBentryALTinterwordstretchfactor}{4}
\providecommand{\BIBentryALTinterwordspacing}{\spaceskip=\fontdimen2\font plus
\BIBentryALTinterwordstretchfactor\fontdimen3\font minus \fontdimen4\font\relax}
\providecommand{\BIBforeignlanguage}[2]{{%
\expandafter\ifx\csname l@#1\endcsname\relax
\typeout{** WARNING: IEEEtran.bst: No hyphenation pattern has been}%
\typeout{** loaded for the language `#1'. Using the pattern for}%
\typeout{** the default language instead.}%
\else
\language=\csname l@#1\endcsname
\fi
#2}}
\providecommand{\BIBdecl}{\relax}
\BIBdecl

\bibitem{EpochNotableModels2024}
\BIBentryALTinterwordspacing
{Epoch AI}, ``Data on notable ai models,'' 2024, accessed: 2024-11-15. [Online]. Available: \url{https://epoch.ai/data/notable-ai-models}
\BIBentrySTDinterwordspacing

\bibitem{EpochMachineLearningHardware2024}
\BIBentryALTinterwordspacing
------, ``Data on machine learning hardware,'' 2024, accessed: 2024-11-15. [Online]. Available: \url{https://epoch.ai/data/machine-learning-hardware}
\BIBentrySTDinterwordspacing

\bibitem{deNearthresholdVoltageDesign2013}
V.~De, ``Near-threshold {{Voltage Design}} in {{Nanoscale CMOS}},'' in \emph{Proceedings of the {{Conference}} on {{Design}}, {{Automation}} and {{Test}} in {{Europe}}}, ser. {{DATE}} '13.\hskip 1em plus 0.5em minus 0.4em\relax San Jose, CA, USA: EDA Consortium, 2013, pp. 612--612.

\bibitem{contiMarsellusHeterogeneousRISCV2024}
F.~Conti, G.~Paulin, A.~Garofalo, D.~Rossi, A.~Di~Mauro, G.~Rutishauser, G.~Ottavi, M.~Eggiman, H.~Okuhara, and L.~Benini, ``Marsellus: {{A Heterogeneous RISC-V AI-IoT End-Node SoC With}} 2--8 b {{DNN Acceleration}} and 30\%-{{Boost Adaptive Body Biasing}},'' \emph{IEEE Journal of Solid-State Circuits}, vol.~59, no.~1, pp. 128--142, Jan. 2024.

\bibitem{nadaliniTOPSRISCVParallel2023}
A.~Nadalini, G.~Rutishauser, A.~Burrello, N.~Bruschi, A.~Garofalo, L.~Benini, F.~Conti, and D.~Rossi, ``A 3 {{TOPS}}/{{W RISC-V Parallel Cluster}} for {{Inference}} of {{Fine-Grain Mixed-Precision Quantized Neural Networks}},'' in \emph{2023 {{IEEE Computer Society Annual Symposium}} on {{VLSI}} ({{ISVLSI}})}, Jun. 2023, pp. 1--6.

\bibitem{prasadSpecializationMeetsFlexibility2023}
A.~Prasad, L.~Benini, and F.~Conti, ``Specialization meets {{Flexibility}}: A {{Heterogeneous Architecture}} for {{High-Efficiency}}, {{High-flexibility AR}}/{{VR Processing}},'' in \emph{Proceedings of the 2023 {{Design Automation Conference}} ({{DAC}} 2023), to Appear}, 2023.

\bibitem{gautschiNearThresholdRISCVCore2017}
M.~Gautschi, P.~D. Schiavone, A.~Traber, I.~Loi, A.~Pullini, D.~Rossi, E.~Flamand, F.~K. G{\"u}rkaynak, and L.~Benini, ``Near-{{Threshold RISC-V Core With DSP Extensions}} for {{Scalable IoT Endpoint Devices}},'' \emph{IEEE Transactions on Very Large Scale Integration (VLSI) Systems}, vol.~25, no.~10, pp. 2700--2713, Oct. 2017.

\bibitem{garofaloXpulpNNEnablingEnergy2021}
A.~Garofalo, G.~Tagliavini, F.~Conti, L.~Benini, and D.~Rossi, ``{{XpulpNN}}: {{Enabling Energy Efficient}} and {{Flexible Inference}} of {{Quantized Neural Networks}} on {{RISC-V Based IoT End Nodes}},'' \emph{IEEE Transactions on Emerging Topics in Computing}, vol.~9, no.~3, pp. 1489--1505, Jul. 2021.

\bibitem{ottaviDustin16CoresParallel2023}
G.~Ottavi, A.~Garofalo, G.~Tagliavini, F.~Conti, A.~D. Mauro, L.~Benini, and D.~Rossi, ``Dustin: {{A}} 16-{{Cores Parallel Ultra-Low-Power Cluster With}} 2b-to-32b {{Fully Flexible Bit-Precision}} and {{Vector Lockstep Execution Mode}},'' \emph{IEEE Transactions on Circuits and Systems I: Regular Papers}, pp. 1--14, 2023.

\bibitem{tortorellaRedMuleMixedprecisionMatrix2023}
Y.~Tortorella, L.~Bertaccini, L.~Benini, D.~Rossi, and F.~Conti, ``{{RedMule}}: {{A}} mixed-precision matrix--matrix operation engine for flexible and energy-efficient on-chip linear algebra and {{TinyML}} training acceleration,'' \emph{Future Generation Computer Systems}, vol. 149, pp. 122--135, Dec. 2023.

\bibitem{garofaloDARKSIDEHeterogeneousRISCV2022}
A.~Garofalo, Y.~Tortorella, M.~Perotti, L.~Valente, A.~Nadalini, L.~Benini, D.~Rossi, and F.~Conti, ``{{DARKSIDE}}: {{A Heterogeneous RISC-V Compute Cluster}} for {{Extreme-Edge On-Chip DNN Inference}} and {{Training}},'' \emph{IEEE Open Journal of the Solid-State Circuits Society}, vol.~2, pp. 231--243, 2022.

\bibitem{garofaloHeterogeneousInMemoryComputing2022}
A.~Garofalo, G.~Ottavi, F.~Conti, G.~Karunaratne, I.~Boybat, L.~Benini, and D.~Rossi, ``A {{Heterogeneous In-Memory Computing Cluster}} for {{Flexible End-to-End Inference}} of {{Real-World Deep Neural Networks}},'' \emph{IEEE Journal on Emerging and Selected Topics in Circuits and Systems}, vol.~12, no.~2, pp. 422--435, Jun. 2022.

\bibitem{pulliniHeterogeneousMultiCoreSystemonChip2017}
A.~Pullini, F.~Conti, D.~Rossi, I.~Loi, M.~Gautschi, and L.~Benini, ``A {{Heterogeneous Multi-Core System-on-Chip}} for {{Energy Efficient Brain Inspired Computing}},'' \emph{IEEE Transactions on Circuits and Systems II: Express Briefs}, vol.~PP, no.~99, pp. 1--1, 2017.

\bibitem{contiIoTEndpointSystemonChip2017}
F.~Conti, R.~Schilling, P.~D. Schiavone, A.~Pullini, D.~Rossi, F.~K. G{\"u}rkaynak, M.~Muehlberghuber, M.~Gautschi, I.~Loi, G.~Haugou, S.~Mangard, and L.~Benini, ``An {{IoT Endpoint System-on-Chip}} for {{Secure}} and {{Energy-Efficient Near-Sensor Analytics}},'' \emph{IEEE Transactions on Circuits and Systems I: Regular Papers}, vol.~64, no.~9, pp. 2481--2494, Sep. 2017.

\bibitem{rossiVegaTenCoreSoC2022}
D.~Rossi, F.~Conti, M.~Eggiman, A.~D. Mauro, G.~Tagliavini, S.~Mach, M.~Guermandi, A.~Pullini, I.~Loi, J.~Chen, E.~Flamand, and L.~Benini, ``Vega: {{A Ten-Core SoC}} for {{IoT Endnodes With DNN Acceleration}} and {{Cognitive Wake-Up From MRAM-Based State-Retentive Sleep Mode}},'' \emph{IEEE Journal of Solid-State Circuits}, vol.~57, no.~1, pp. 127--139, Jan. 2022.

\bibitem{prasadSiracusa16Nm2024}
A.~S. Prasad, M.~Scherer, F.~Conti, D.~Rossi, A.~D. Mauro, M.~Eggimann, J.~T. G{\'o}mez, Z.~Li, S.~S. Sarwar, Z.~Wang, B.~D. Salvo, and L.~Benini, ``Siracusa: {{A}} 16 nm {{Heterogenous RISC-V SoC}} for {{Extended Reality With At-MRAM Neural Engine}},'' \emph{IEEE Journal of Solid-State Circuits}, pp. 1--15, 2024.

\bibitem{paulinOccamy432Core2812024}
G.~Paulin, P.~Scheffler, T.~Benz, M.~Cavalcante, T.~Fischer, M.~Eggimann, Y.~Zhang, N.~Wistoff, L.~Bertaccini, L.~Colagrande, G.~Ottavi, F.~K. G{\"u}rkaynak, D.~Rossi, and L.~Benini, ``Occamy: {{A}} 432-{{Core}} 28.1 {{DP-GFLOP}}/s/{{W}} 83\% {{FPU Utilization Dual-Chiplet}}, {{Dual-HBM2E RISC-V-Based Accelerator}} for {{Stencil}} and {{Sparse Linear Algebra Computations}} with 8-to-64-bit {{Floating-Point Support}} in 12nm {{FinFET}},'' in \emph{2024 {{IEEE Symposium}} on {{VLSI Technology}} and {{Circuits}} ({{VLSI Technology}} and {{Circuits}})}, Jun. 2024, pp. 1--2.

\bibitem{miro-panadesSamurAIVersatileIoT2022}
I.~{Miro-Panades}, B.~Tain, J.-F. Christmann, D.~Coriat, R.~Lemaire, C.~Jany, B.~Martineau, F.~Chaix, G.~Waltener, E.~Pluchart, J.-P. Noel, A.~Makosiej, M.~Montoya, S.~{Bacles-Min}, D.~Briand, J.-M. Philippe, Y.~Thonnart, A.~Valentian, F.~Heitzmann, and F.~Clermidy, ``{{SamurAI}}: {{A Versatile IoT Node With Event-Driven Wake-Up}} and {{Embedded ML Acceleration}},'' \emph{IEEE Journal of Solid-State Circuits}, pp. 1--0, 2022.

\bibitem{juSystolicNeuralCPU2023}
Y.~Ju and J.~Gu, ``A {{Systolic Neural CPU Processor Combining Deep Learning}} and {{General-Purpose Computing With Enhanced Data Locality}} and {{End-to-End Performance}},'' \emph{IEEE Journal of Solid-State Circuits}, vol.~58, no.~1, pp. 216--226, Jan. 2023.

\bibitem{jainTinyVersTinyVersatile2023}
V.~Jain, S.~Giraldo, J.~D. Roose, L.~Mei, B.~Boons, and M.~Verhelst, ``{{TinyVers}}: {{A Tiny Versatile System-on-Chip With State-Retentive eMRAM}} for {{ML Inference}} at the {{Extreme Edge}},'' \emph{IEEE Journal of Solid-State Circuits}, pp. 1--12, 2023.

\bibitem{houshmandDIANAEndtoEndHybrid2023}
P.~Houshmand, G.~M. Sarda, V.~Jain, K.~Ueyoshi, I.~A. Papistas, M.~Shi, Q.~Zheng, D.~Bhattacharjee, A.~Mallik, P.~Debacker, D.~Verkest, and M.~Verhelst, ``{{DIANA}}: {{An End-to-End Hybrid DIgital}} and {{ANAlog Neural Network SoC}} for the {{Edge}},'' \emph{IEEE Journal of Solid-State Circuits}, vol.~58, no.~1, pp. 203--215, Jan. 2023.

\bibitem{moonsEnvision26to10TOPSSubwordparallel2017}
B.~Moons, R.~Uytterhoeven, W.~Dehaene, and M.~Verhelst, ``Envision: {{A}} 0.26-to-{{10TOPS}}/{{W}} subword-parallel dynamic-voltage-accuracy-frequency-scalable {{Convolutional Neural Network}} processor in 28nm {{FDSOI}},'' in \emph{2017 {{IEEE International Solid-State Circuits Conference}} ({{ISSCC}})}, Feb. 2017, pp. 246--247.

\bibitem{desoli9TOPSDeepConvolutional2017}
G.~Desoli, N.~Chawla, T.~Boesch, S.~p~Singh, E.~Guidetti, F.~D. Ambroggi, T.~Majo, P.~Zambotti, M.~Ayodhyawasi, H.~Singh, and N.~Aggarwal, ``A 2.{{9TOPS}}/{{W}} deep convolutional neural network {{SoC}} in {{FD-SOI}} 28nm for intelligent embedded systems,'' in \emph{2017 {{IEEE International Solid-State Circuits Conference}} ({{ISSCC}})}, Feb. 2017, pp. 238--239.

\bibitem{tambe2212nm182023}
T.~Tambe, J.~Zhang, C.~Hooper, T.~Jia, P.~N. Whatmough, J.~Zuckerman, M.~C.~D. Santos, E.~J. Loscalzo, D.~Giri, K.~Shepard, L.~Carloni, A.~Rush, D.~Brooks, and G.-Y. Wei, ``22.9 {{A}} 12nm 18.{{1TFLOPs}}/{{W Sparse Transformer Processor}} with {{Entropy-Based Early Exit}}, {{Mixed-Precision Predication}} and {{Fine-Grained Power Management}},'' in \emph{2023 {{IEEE International Solid- State Circuits Conference}} ({{ISSCC}})}.\hskip 1em plus 0.5em minus 0.4em\relax San Francisco, CA, USA: IEEE, Feb. 2023, pp. 342--344.

\bibitem{ditzelAcceleratingMLRecommendation2022}
D.~R. Ditzel and t.~E. {team}, ``Accelerating {{ML Recommendation With Over}} 1,000 {{RISC-V}}/{{Tensor Processors}} on {{Esperanto}}'s {{ET-SoC-1 Chip}},'' \emph{IEEE Micro}, vol.~42, no.~3, pp. 31--38, May 2022.

\bibitem{mo12TOPSQuantized2022}
H.~Mo, W.~Zhu, W.~Hu, Q.~Li, A.~Li, S.~Yin, S.~Wei, and L.~Liu, ``A 12.1 {{TOPS}}/{{W Quantized Network Acceleration Processor With Effective-Weight-Based Convolution}} and {{Error-Compensation-Based Prediction}},'' \emph{IEEE Journal of Solid-State Circuits}, vol.~57, no.~5, pp. 1542--1557, May 2022.

\bibitem{hager113MetisAIPU2024}
P.~A. Hager, B.~Moons, S.~Cosemans, I.~A. Papistas, B.~Rooseleer, J.~V. Loon, R.~Uytterhoeven, F.~Zaruba, S.~Koumousi, M.~Stanisavljevic, S.~Mach, S.~Mutsaards, R.~K. Aljameh, G.~H. Khov, B.~Machiels, C.~Olar, A.~Psarras, S.~Geursen, J.~Vermeeren, Y.~Lu, A.~Maringanti, D.~Ameta, L.~Katselas, N.~H{\"u}tter, M.~Schmuck, S.~Sivadas, K.~Sharma, M.~Oliveira, R.~Aerne, N.~Sharma, T.~Soni, B.~Bussolino, D.~Pesut, M.~Pallaro, A.~Podlesnii, A.~Lyrakis, Y.~Ruffiner, M.~Dazzi, J.~Thiele, K.~Goetschalckx, N.~Bruschi, J.~Doevenspeck, B.~Verhoef, S.~Linz, G.~Garcea, J.~Ferguson, I.~Koltsidas, and E.~Eleftheriou, ``11.3 {{Metis AIPU}}: {{A}} 12nm {{15TOPS}}/{{W}} 209.{{6TOPS SoC}} for {{Cost-}} and {{Energy-Efficient Inference}} at the {{Edge}},'' in \emph{2024 {{IEEE International Solid-State Circuits Conference}} ({{ISSCC}})}, vol.~67, Feb. 2024, pp. 212--214.

\bibitem{yiRDCIMRISCVSupported2024}
W.~Yi, K.~Mo, W.~Wang, Y.~Zhou, Y.~Zeng, Z.~Yuan, B.~Cheng, and B.~Pan, ``{{RDCIM}}: {{RISC-V Supported Full-Digital Computing-in-Memory Processor With High Energy Efficiency}} and {{Low Area Overhead}},'' \emph{IEEE Transactions on Circuits and Systems I: Regular Papers}, vol.~71, no.~4, pp. 1719--1732, Apr. 2024.

\bibitem{desoli16740310TOPSSRAMBased2023}
G.~Desoli, N.~Chawla, T.~Boesch, M.~Avodhyawasi, H.~Rawat, H.~Chawla, {\relax VS}.~Abhijith, P.~Zambotti, A.~Sharma, C.~Cappetta, M.~Rossi, A.~De~Vita, and F.~Girardi, ``16.7 {{A}} 40-{{310TOPS}}/{{W SRAM-Based All-Digital Up}} to 4b {{In-Memory Computing Multi-Tiled NN Accelerator}} in {{FD-SOI}} 18nm for {{Deep-Learning Edge Applications}},'' in \emph{2023 {{IEEE International Solid-State Circuits Conference}} ({{ISSCC}})}, Feb. 2023, pp. 260--262.

\bibitem{chenTVMAutomatedEndtoEnd2018}
T.~Chen, T.~Moreau, Z.~Jiang, L.~Zheng, E.~Yan, H.~Shen, M.~Cowan, L.~Wang, Y.~Hu, L.~Ceze, C.~Guestrin, and A.~Krishnamurthy, ``{{TVM}}: {{An Automated End-to-End Optimizing Compiler}} for {{Deep Learning}},'' in \emph{13th {{USENIX Symposium}} on {{Operating Systems Design}} and {{Implementation}} ({{OSDI}} 18)}, 2018, pp. 578--594.

\bibitem{burrelloDORYAutomaticEndtoEnd2021}
A.~Burrello, A.~Garofalo, N.~Bruschi, G.~Tagliavini, D.~Rossi, and F.~Conti, ``{{DORY}}: {{Automatic End-to-End Deployment}} of {{Real-World DNNs}} on {{Low-Cost IoT MCUs}},'' \emph{IEEE Transactions on Computers}, vol.~70, no.~8, pp. 1253--1268, Aug. 2021.

\bibitem{schererDeeployEnablingEnergyEfficient2024}
M.~Scherer, L.~Macan, V.~J.~B. Jung, P.~Wiese, L.~Bompani, A.~Burrello, F.~Conti, and L.~Benini, ``Deeploy: {{Enabling Energy-Efficient Deployment}} of {{Small Language Models}} on {{Heterogeneous Microcontrollers}},'' \emph{IEEE Transactions on Computer-Aided Design of Integrated Circuits and Systems}, vol.~43, no.~11, pp. 4009--4020, Nov. 2024.

\bibitem{EdgeAIVision2023}
\BIBentryALTinterwordspacing
{Edge AI Vision Alliance}, ``Data on notable ai models,'' 2023, accessed: 2024-12-23. [Online]. Available: \url{https://www.edge-ai-vision.com/2023/07/average-design-cost-for-advanced-performance-multicore-socs-to-reach-12-2m-by-2027/}
\BIBentrySTDinterwordspacing

\end{thebibliography}
\end{document}